\documentclass[a4paper]{jpconf}

\usepackage{citesort}

\def\mmm{\mathcal{V}}

\def\pertp{\varepsilon}

\def\gfam{\hat g}
\def\Supfam{\hat \Sigma}

\def\hfam{\hat h}
\def\kfam{\hat \kappa}

\def\gfamp{g}
\def\Supfamp{\Sigma}
\def\hfamp{h}
\def\kfamp{\kappa}

\def\gback{g}
\def\fpt{K_1}
\def\spt{K_2}

\def \tp {t_+}
\def \phip {\varphi_+}
\def \rrp {r_+}
\def \thetap {\theta_+}
\def \tm {t_-}
\def \phim {\varphi_-}
\def \rrm {r_-}
\def \thetam {\theta_-}

\def\ro{a} 

\def\oppert{\omega'}
\def\topert{(\omega-\Omega)}

\def\defi{:=}

\def\Qt2{\hat{Q}_2}

\def\energy{E}
\def\pressure{P}
\def\Eb{\energy}
\def\Ep{\energy^{(1)}}
\def\Epp{\energy^{(2)}}
\def\Eppz{\energy^{(2)}_0}
\def\Eppt{\energy^{(2)}_2}
\def\Pb{\pressure}
\def\Pp{\pressure^{(1)}}
\def\Ppp{\pressure^{(2)}}
\def\Pppz{\pressure^{(2)}_0}
\def\Pppt{\pressure^{(2)}_2}

\def\pstarHz{p_0^*}
\def\pressurestar{\mathcal{P}}

\def\Pspp{\pressurestar}
\def\Psppz{\Pspp_0}

\def \hk {h_0}

\def \mk {m_0}

\def \kkt {k_2}

\def \hkt {h_2}

\def \mkt {m_2}

\usepackage{graphicx}
\begin{document}
\title{Hartle's model within the general theory of
perturbative matchings: the change in mass}

\author{Borja Reina, Ra\"ul Vera}

\address{Dept. of Theoretical Physics and History of Science,\\ University of the Basque Country UPV/EHU,\\
644 PK, Bilbao 48080, Basque Country, Spain}

\ead{borja.reina@ehu.es, raul.vera@ehu.es}

\begin{abstract}
Hartle's model provides the most widely used analytic framework
to describe isolated compact bodies rotating slowly in equilibrium up to second order in
perturbations in the context of General Relativity. Apart from some explicit
assumptions, there are some implicit, like the ``continuity'' of
the functions in the perturbed metric across the surface of the body.
In this work we sketch the basics for the analysis of the second
order problem using the modern theory of perturbed matchings. In
particular, the result we present is that when the
energy density of the fluid in the static configuration does not vanish at
the boundary, one of the functions of the second order perturbation
in the setting of the original work by Hartle is not continuous.
This discrepancy affects the calculation of the change in mass
of the rotating star with respect to the static configuration
needed to keep the central energy density unchanged.
\end{abstract}

\section{Introduction}
One of the aims of our programme to study Hartle's model \cite{Hartle1967} within the general modern theory of
perturbations is to analyse with rigour the implicit assumptions made in the original
model regarding the continuity and/or differentiability of the
functions in the metric of the perturbed configuration at the boundary of the star.
In order to do that we have studied the problem of the perturbed matching by making use
of modern spacetime and matching perturbation theory, only achieved in full generality
and to second order in \cite{Mars2005}.
This work is a parallel continuation of the analysis started in \cite{ReinaVera_ERE2012}
based on the theoretical analysis of a completely general
perturbative approach to second order around static configurations of the exterior
(asymptotically flat) vacuum problem of stationary and axisymmetric bodies with
arbitrary matter content presented in \cite{MaMaVe2007}.

In this short contribution to the ERE2014 we present the main result concerning
the matching problem and continuity of functions in Hartle's model,
which eventually implies a correction of the change in mass of the star
needed by the perturbed configuration to keep the value of the central energy density
unchanged with respect to the static configuration.

\subsection{Hartle's model}
Hartle's model \cite{Hartle1967} describes the equilibrium configuration of a slowly rotating isolated body in a perturbation scheme in the context of General Relativity. It is a global model, where the interior of the body is a perfect fluid with a barotropic equation of state, no convective motions and rigid rotation and the exterior is an asymptotically flat vacuum. These are matched across a timelike hypersurface that represents the common boundary between the star and the vacuum.

The whole model is stationary and axially symmetric, but also equatorial symmetry is assumed. The perturbations are computed up to second order around a static, spherically symmetric configuration and they are driven by a single parameter. In this section we present the basics
of Hartle's model and the relevant results for the present work just as they are presented in the original paper
\cite{Hartle1967}, except that we make explicit the appearance of the perturbation parameter, denoted here
as $\pertp$. In particular, $\Omega$ in \cite{Hartle1967} appears here as $\pertp\Omega$
for some constant $\Omega$.

The ``perturbed'' metric used in \cite{Hartle1967} describes both the interior and exterior regions at once 
\begin{eqnarray}
\gfamp_{\pertp} &=& -e^{\nu(r)}\left(1+2 \pertp^2 h(r,\theta)\right)dt^2 + e^{\lambda(r)}\left(1+ 2 \pertp^2 \frac{e^{\lambda(r)}m(r,\theta)}{r} \right)dr^2 \nonumber\\
&& + r^2(1+2 \pertp^2 k(r,\theta))\left[d\theta^2 + \sin ^2 \theta (d\varphi-\pertp \omega(r,\theta) dt)^2\right]\,
+ \mathcal{O}(\pertp^3), \label{gefamily_Hartle}
\end{eqnarray}
where $r$ runs from 0 to $\infty$.
It is \emph{implicitly assumed} that in this set of coordinates this metric is, at least, continuous. 
The static and spherically symmetric background configuration is determined by the functions $\nu(r)$ and $\lambda(r)$. In this background, the common boundary of the interior and the exterior is located at $r=\ro$, so that the fluid region extends up to $r=\ro$ and the vacuum from there to $\infty$.
The first order perturbation is described by the function $\omega (r, \theta)$ and the second order perturbation by $h(r,\theta)$, $m(r,\theta)$ and $k(r,\theta)$.
As it is well known, the choice of the perturbation parameter $\pertp$ is not relevant, 
since one can obtain other families of solutions by scaling.
The physics of the model will restrict the scalability (see Eq. (1) in \cite{Hartle1967}).
Therefore, the functions in (\ref{gefamily_Hartle}) can be taken to
correspond to the functions with the same name in \cite{Hartle1967}.

The energy density $\Eb$ and pressure  $\Pb$  of the background interior read from
\begin{equation}
\lambda' = \frac{1}{r}(1-e^\lambda)+re^\lambda 8\pi\Eb,
\quad
\nu' =\frac{1}{r}(e^\lambda-1)+re^\lambda 8\pi\Pb \label{eq:nuprime}
\end{equation}
for the fluid with $\vec u=e^{-\nu/2}\partial_{t}$.
These are combined with the pressure isotropy condition to give the usual TOV equation, which 
can be integrated given a value of the central energy density.
It is useful to define $M(r):= r(1-e^{-\lambda})/2$ and $j(r):=e^{-(\lambda + \nu)/2}$ in order to cast the equations for the perturbations in a compact way. In vacuum, $j(r)=1$ and $M(r)=M$, and thus
\begin{equation}
e^{-\lambda(r)} = e^{\nu(r)} = 1-\frac{2M}{r},
\label{back_vacuum}
\end{equation}
where $M$ is a constant recognized as the mass of the nonrotating star by the assumption of continuity of $\lambda(r)$ at $r=\ro$.
The first order perturbation for the fluid velocity reads $\vec{u}^{(1)} = \pertp \Omega \partial_\varphi$, for some constant $\Omega$, and the field equations ($t\varphi$) provide a PDE for $\omega (r, \theta)$. At this point,  $\omega$ is \emph{implicitly} assumed to be $\mathcal{C}^1$,
so that regularity conditions at the origin together with asymptotic flatness are used to show that $\omega$ must be a function of $r$ alone. Thence, $\omega(r)$ is integrated given its value at the origin.
In vacuum  $\omega(r) = 2Jr^{-3}$ for some constant $J$, which can be
determined from the interior by continuity at $r=\ro$ \cite{Hartle1967}.

At second order the fluid flow is proportional to $\partial_t$, and the dependence of $\omega$ propagates through the field equations leading to the  finite expansions
\begin{equation}
h(r,\theta) = \hk(r) + \hkt(r) P_2(\cos \theta), \, m(r,\theta) = \mk(r) + \mkt(r) P_2(\cos \theta),\, k(r,\theta) = \kkt(r) P_2(\cos \theta),\label{angularexpansions}
\end{equation}
\emph{where $r$ is chosen} so that $k_0(r)=0$. That corresponds to a
choice of gauge at second order, which we call $k$-gauge. The problems
for the $l=0$ and $l=2$ sectors separate and are studied
independently. In this text we restrict ourselves to the $l=0$ problem
(for a complete analysis see \cite{Hartle1967,ReinaVera2014}). The
perfect fluid equations for the set $\{h_0,m_0\}$ are presented in
\cite{Hartle1967}, instead, in terms of the set $\{\pstarHz,m_0\}$,
where $\pstarHz$ (called the pressure perturbation factor in
\cite{Hartle1967}) is related to $h_0$ by 
\begin{equation}
\gamma = h_0(r) + \pstarHz (r) -\frac{1}{3}r^2e^{-\nu(r)}(\omega(r) - \Omega)^2, \label{eq:hydro}
\hspace{-2mm}
\end{equation}
for some constant $\gamma$. The equations are written in  \cite{Hartle1967} in terms of
another radial coordinate $R$, which runs from 0 to $\ro$. However, at this point 
we can treat $R$ as a  dummy variable, which for convenience we recall here $r (\in(0,\ro])$.
The system for $\{\pstarHz(r),m_0(r)\}$ in $r\in(0,\ro]$  reads 
\begin{eqnarray}
&&\pstarHz{}' =\label{eq:Ppf0}\\
&&  \frac{r^4 j^2}{12(r-2M(r))}\oppert^2 + \frac{1}{3} \left(\frac{r^3j^2 \topert^2}{r-2M(r)} \right)' -\frac{4\pi (\Eb+\Pb)r^2}{r-2 M(r)}\pstarHz
- \frac{m_0 r^2 }{(r-2M(r))^2}\left( 8\pi \Pb + \frac{1}{r^2}\right) \nonumber\\
&&m_0'=4\pi r^2 \frac{\Eb'}{\Pb'}(\Eb+\Pb)\pstarHz +\frac{1}{12}j^2r^4\oppert^2-\frac{2}{3}r^3 jj'\topert^2, \label{eq:m0}
\end{eqnarray}
where $'$ denotes a derivative with respect to the argument ($r$ in this case).
The system is solved imposing regularity conditions at the origin, plus $\pstarHz(0)= 0$, thus
forcing the central energy density to stay unchanged with respect to the static configuration \cite{Hartle1967}.
Therefore, $\gamma$ takes the value $h_0(0)$,
which in principle is arbitrary ($h_0$ is determined up to an additive constant).
In vacuum, the equations for $m_0$ and $h_0$ provide the solutions for $r\geq\ro$
\begin{equation}
 m_0(r) =  \delta M - \frac{J^2}{r^3},\quad h_0 (r)= -\frac{\delta M}{r-2M}+\frac{J^2}{r^3(r -2M)}, \label{sol:m0ext}
\end{equation}
where $\delta M$ is a constant and the free additive constant in $h_0$ has been set so that $h_0$ vanishes at infinity. $\delta M$ is recognized in the asymptotic region as the
change in mass needed to keep the central energy density unchanged.
The assumed continuity of $m_0$ at $r=\ro$ in \cite{Hartle1967} yields
\begin{equation}
\delta M = m_0(\ro) + \frac{J^2}{\ro^3},
\label{deltaM_Hartle}
\end{equation}
which determines $\delta M$ in terms of the interior quantities, since
$m_0(\ro)$ is the value of the function $m_0$ obtained by the integration of the system
(\ref{eq:Ppf0}), (\ref{eq:m0}) out from the origin.
Continuity of $h_0(r)$ is  used to fix the freedom left in $h_0$ in the interior and thus
fix $\gamma$.

\section{The perturbation method} \label{section:method}
The modern view of perturbation theory starts with a
family of spacetimes $(\mmm_{\pertp},\gfam_\pertp)$ with diffeomorphically related manifolds,
from where a background spacetime say $(\mmm_0,\gback)$ is singled out, so that
$\mmm_0\defi \mmm_{\pertp=0}$ and $\gback\defi \gfam_{\pertp=0}$.
The diffeomorphism $\psi_\pertp:\mmm_0\to\mmm_\pertp$ used to
identify points of the manifolds is employed to
pull back $\gfam_\pertp$ onto the background spacetime, $\gfamp_\pertp\defi \psi^*_\pertp(\gfam_\pertp)$.
The family of tensors $\gfamp_\pertp$ 
defines the basic perturbative scheme on $(\mmm_0,\gback)$, where
$\gback\defi \gfam_{\pertp=0}=\gfamp_{\pertp=0} $, in the particular gauge defined by $\psi_\pertp$.
The metric perturbation tensors are now simply defined as the derivatives  of $\gfamp_\pertp$
with respect to $\pertp$ at $\pertp=0$ at each order of derivation.
$\fpt\defi\partial_\pertp \gfamp_\pertp|_{\pertp=0}$ and $\spt\defi\partial_\pertp\partial_\pertp \gfamp_\pertp|_{\pertp=0}$ will refer to the first and second
metric perturbation tensors.

Given the above family (\ref{gefamily_Hartle}),
the spherically symmetric and static background metric reads
\begin{equation}
\gback= -e^{\nu(r)} d{t}^ 2 + e^{\lambda(r)} d{r}^ 2+{r}^ 2(d{\theta}^2+\sin^2 \theta d {\varphi}^ 2),\label{eq:g0}
\end{equation}
while 
the perturbation tensors to second order take the form
$\fpt  = 
 -2r^2\, \omega(r,\theta) \sin ^2 \theta dt d\varphi$ and
\begin{eqnarray}
\spt &=& \left(-4 e^{\nu(r)} h(r, \theta) + 2r^2  {\omega}^2(r, \theta)\sin ^2 \theta\right)dt^2 + 4 \frac{e^{2\lambda(r)}}{r} m(r, \theta) dr^2
\nonumber\\
&&
+4 r^2 k(r, \theta)
(d\theta^2 + \sin ^2 \theta d\varphi^2).\label{sopert_tensor}
\end{eqnarray}

Matter fields are  introduced as an $\pertp$-family of tensors $T_\pertp$
on  $(\mmm_0,g)$, with corresponding perturbations defined again by taking $\pertp$-derivatives at $\pertp=0$. For an energy momentum tensor of the form 
$
T_{\pertp} = (\energy_\pertp + \pressure_\pertp) u_\pertp \otimes u_\pertp + \pressure_\pertp g_{\pertp{}},
$
those will be given in terms of the corresponding (unit) fluid flow $u_\pertp$ energy density  $\energy_\pertp$ and pressure $\pressure_\pertp$,
expanded as $\vec u_\pertp=\vec u+\pertp \vec u^{(1)}+\frac{1}{2}\pertp^2\vec u^{(2)}
+\mathcal{O}(\pertp^3)$,
\begin{equation}
\energy_\pertp = \Eb + \pertp \Ep + \frac{1}{2}\pertp^2 \Epp + \mathcal{O}(\pertp^3),\quad \pressure_\pertp = \Pb + \pertp \Pp + \frac{1}{2}\pertp^2 \Ppp+  \mathcal{O}(\pertp^3). \label{Pressure_epsilon}
\end{equation}
The Einstein's field equations for 
$g_{\pertp}$ read $G(g_\pertp)_{\alpha\beta}=8\pi T_\pertp{}_{\alpha\beta}$,
from where the first and second order field equations are obtained as
$\partial_\pertp G(g_\pertp)_{\alpha\beta}|_{\pertp=0}
= 8 \pi \partial_\pertp T_\pertp{}_{\alpha\beta}|_{\pertp=0}$ and 
$\partial_\pertp \partial_\pertp G(g_\pertp)_{\alpha\beta}|_{\pertp=0}
= 8 \pi \partial_\pertp \partial_\pertp T_\pertp{}_{\alpha\beta}|_{\pertp=0}$, respectively.

In order to construct the global model we consider a perfect fluid interior ($+$)
region to be matched to an asymptotically flat vacuum exterior ($-$) region.
In the exact case, the matching of spacetimes with boundary, say $(\mmm^\pm,g^\pm,\Sigma^\pm)$
requires an identification of the boundaries, $\Sigma^+$ and $\Sigma^-$,
that 
defines an abstract manifold $\Sigma$ diffeomorphic to both.
The matching conditions demand the existence of one
such identification for which the first and second fundamental forms
as computed from either side, $\hfamp^\pm$ and $\kfamp^\pm$, agree.
Let us assume that this construction holds for each pair
$(\mmm^\pm_\pertp,\gfam^\pm_\pertp,\Supfam^\pm_\pertp)$
(with $\varepsilon$ fixed), so that there exists a family of diffeomorphically related
hypersurfaces $\Supfam_\pertp$ 
on which $\hfam^+_\pertp=\hfam^-_\pertp$,
$\kfam^+_\pertp=\kfam^-_\pertp$ hold.
By construction $\Sigma_0=\Supfam_{\pertp=0}$ is
the matching hypersurface  of the already matched background.
These conditions are formulated as an $\pertp$-family of equations
defined on $\Sigma_0$ as follows. The spacetime gauges $\psi^\pm_\pertp$
project (down or up \cite{Mars2005}) the $\Supfam^\pm_\pertp$ families to the background
spacetime $(\mathcal{V}^\pm_0, g^\pm)$ giving rise to corresponding families of
hypersurfaces $\Supfamp^\pm_\pertp$ in $(\mathcal{V}^\pm_0, g^\pm)$.
In general, $\Supfamp^\pm_\pertp$ do not coincide with  $\Supfamp^\pm_0$.
The prescription of how
points in $\Sigma_0^\pm$ map to $\Supfamp^\pm_\pertp$
is done through yet another diffeomorphism $\Phi_\pertp:\Sigma_0\to \Supfam_\pertp $ that identifies $\Supfam_\pertp$ pointwise. This defines a second
gauge freedom, the \emph{hypersurface gauge}.
The combination of the two gauges
allows us (i) to construct the mapping of the
families $\Supfamp^\pm_\pertp$ to $\Supfamp^\pm_0$ at each side,
which describes the pointwise
deformation of the hypersurfaces with respect to $\Supfamp^\pm_0(=\Supfamp_0)$
in the gauges defined by $\psi^\pm_\pertp$ at either side, and
(ii) to pull back $\hfam^\pm_\pertp$ and $\kfam^\pm_\pertp$,
down (or up) to $(\Sigma_0,\hfamp)$ with $\hfamp=\hfam^+_0=\hfamp^-_0$,
and thus obtain the families of tensors $\hfamp^\pm_\pertp$ and $\kfamp^\pm_\pertp$
on $(\Sigma_0,\hfamp)$ \cite{Mars2005}.
The $\pertp$-derivatives evaluated on $\pertp$
define $\hfamp^{(1)}{}^\pm$, $\hfamp^{(2)}{}^\pm$, $\kfamp^{(1)}{}^\pm$, and $\kfamp^{(2)}{}^\pm$. On the other hand, the first and second order
of the deformation is encoded in the 
the first and second  order deformation vectors \cite{Mars2005}
\begin{equation}
\vec{Z}^\pm_1=Q^\pm_1 \vec n^\pm + \vec T^\pm_1,\qquad
\vec{Z}^\pm_2=Q^\pm_2 \vec n^\pm + \vec T^\pm_2,
\label{eq:Zetas0}
\end{equation}
where $\vec n^\pm$ are the unit normals to $\Supfamp_0^\pm$,
$Q^\pm$ are normal components and $\vec T^\pm$ the tangent parts to $\Sigma_0$.
Apart from the background configuration and the tensors $\fpt^\pm$ and $\spt^\pm$,
$\vec{Z}^\pm_1$ and $\vec{Z}^\pm_2$ are (unknown) ingredients needed to compute
$\hfamp^{(1)}{}^\pm$, $\hfamp^{(2)}{}^\pm$, $\kfamp^{(1)}{}^\pm$, and $\kfamp^{(2)}{}^\pm$.
The explicit expressions are found in Propositions 2 and 3 in \cite{Mars2005}
(see also \cite{BattyeCarter,Mukohyama00} for the first order case).
The
perturbed matching conditions to first and second order are
\begin{equation}
\hfamp^{(1)}{}^+=\hfamp^{(1)}{}^-,\quad
\kfamp^{(1)}{}^+=\kfamp^{(1)}{}^-,\qquad
\hfamp^{(2)}{}^+=\hfamp^{(2)}{}^-,\quad
\kfamp^{(2)}{}^+=\kfamp^{(2)}{}^-.
\label{pertmatch}
\end{equation}
The tensors $\hfamp^{(1)}{}^\pm$, $\hfamp^{(2)}{}^\pm$,
$\kfamp^{(1)}{}^\pm$, and $\kfamp^{(2)}{}^\pm$ are, by construction,
spacetime gauge invariant (they are defined on $(\Sigma_0,\hfamp)$),
but not hypersurface gauge invariant \cite{Mars2005}.
However, \emph{the equations} (\ref{pertmatch})
are both spacetime and hypersurface gauge invariant
\cite{Mars2005}.
Fulfilling the matching conditions at each order requires showing the existence
of two vectors $\vec Z^\pm$ (at each order) such that the equations are satisfied.

The vectors $\vec Z$ depend fully on both spacetime and hypersurface gauges.
Both ($\pm$) can be set to zero simultaneously using the appropriate
spacetime gauges $\psi_\pertp$ at each side. But one has to be careful, then. 
Let us remark the fact that generally in the literature on perturbed matching
it has been common to use spacetime gauges for which $\vec Z^\pm_{1/2}=0$,
the so-called ``co-moving'' gauges. This is fine if no other condition
has been used to restrict the form in which the families $g_\pertp^\pm$
are written, or equivalently, the form of $\fpt^\pm$ and $\spt^\pm$.
But this is precisely the case in using the $k$-gauge in (\ref{gefamily_Hartle}).
In order to compute the perturbed matching conditions in the $k$-gauge we need
to consider the deformation vectors $\vec Z$, which, in fact, will determine
how the surface gets perturbed form the point of view of the $k$-gauge.

\section{Background configuration}

The interior and exterior regions are described, respectively, by 
a couple of spacetimes $(\mathcal{V}^+_0,\gback^+)$ and $(\mathcal{V}^-_0,\gback^-)$ with corresponding boundaries
$\Sigma_0^+$ and $\Sigma_0^-$
and coordinates $\{t_\pm,r_\pm,\theta_\pm,\varphi_\pm\}$, so that
(\ref{eq:g0}) holds with $+$ and $-$. Due to the symmetries the
hypersurfaces $\Sigma_0^+$ and $\Sigma_0^-$
can be finally cast as (see e.g. \cite{Mars2007})
$\Sigma_0^+ = \{\tp= \tau, \rrp=\ro , \phip = \phi, \thetap = \vartheta \}$,
$\Sigma_0^- = \{\tm= \tau, \rrm=\ro , \phim = \phi, \thetam = \vartheta \}$
without loss of generality. The coordinates $\{\tau,\phi,\vartheta\}$ parametrize
$\Sigma_0 (\equiv \Sigma_0^+= \Sigma_0^-)$.
The unit normals to $\Sigma_0^\pm$ are taken to be $\vec{n}^\pm = -e^{-\frac{\lambda_\pm(\ro)}{2}}\partial_{r^\pm}|_{\Sigma^\pm_0}$, under the condition that $\vec n^+$ points $\mathcal{V}_0^+$ inwards
and $\vec n^-$ points $\mathcal{V}_0^-$ outwards.
This convention follows in order to call $\mathcal{V}_0^+$ the \emph{interior} and
$\mathcal{V}_0^-$ the \emph{exterior}.

The matching conditions are found to be equivalent to the set of equations (see e.g. \cite{Mars2007})
\begin{equation}
[\nu]=0,\quad [\nu']=0,\quad [\lambda]=0, \label{background_matching}
\end{equation}
where 
$[f]=f^+|_{\Sigma^+_0}-f^-|_{\Sigma^-_0}$ for objects $f^\pm$
defined at either side, and a prime denotes differentiation with
respect to the corresponding radial coordinate, i.e. $\rrp$ or $\rrm$ accordingly.
Given $f^\pm$ satisfying $[f]=0$, we denote by $f|_{\Sigma_0}$ either of the equivalent $f^+|_{\Sigma^+_0}$
or $f^-|_{\Sigma^+_0}$.

The perfect fluid of the static background interior ($+$) is already described above,
with  $\Eb$ and $\Pb$ given by (\ref{eq:nuprime}), while in the vacuum region ($-$) we have
(\ref{back_vacuum}) for $\lambda_-(r_-)$ and $\nu_-(r_-)$.
The matching conditions (\ref{background_matching}) imply, in particular, that
$
\nu_+(\ro)=-\lambda_+(\ro)=\log\left(1-\frac{2M}{\ro}\right).
$
The expressions for
the differences of the derivative of the functions of the metric
in terms of the differences of the fluid variables on $\Supfamp_0$ are thus found to be
\begin{equation}
\left[\nu'\right] = \ro e^{\lambda(a)}8\pi\Pb(\ro)=0, \quad \left[\lambda'\right]  =  \ro e^{\lambda(\ro)} 8\pi \Eb(\ro), \quad
\left[\nu''\right] = \left(1+\frac{\ro\nu'(\ro)}{2}\right)e^{\lambda(\ro)}8\pi\Eb(\ro).\label{nuplambdapnupp}
\end{equation}

Let us remark that whereas the matching conditions imply
that $\Pb(r_+)$ must vanish on $\Sigma_0$, the energy density
$\Eb(\ro)$ stays free. Its value will be determined, if any,
by the barotropic equation of state (EOS). In most cases the typical
EOS's for neutron stars show a behaviour in which the pressure and
energy density vanish together. This is also the case of the
polytropic EOS, but one can consider many other cases where this
behaviour no longer stands. An example is a fluid with constant density \cite{Chandra_Miller1974},
and a more realistic case, strange quark matter EOS's \cite{ColpiMiller}.

\section{Second order perturbations}
The study of the first order perturbations is out of the scope of this paper, but as a summary, we find that one can set $[\omega]= 0$, whereas $[\omega']=0$ always holds. It can then be proven that 
the first order configuration described in \cite{Hartle1967} (see above) follows (see \cite{ReinaVera_ERE2012}).
Also, $\Ep=\Pp=0.$

At this point we assume that the functions $\{m^\pm,k^\pm,h^\pm\}$
in both regions satisfy (\ref{angularexpansions}).
The second order field equations
for the perfect fluid, whose second order perturbation vector is $\vec{u}^{(2)}= e^{-3\nu/2}\left\{\Omega^2 g_{\varphi\varphi}+2\Omega \fpt{}_{t\varphi} +\spt{}_{tt}/2  \right\} \partial_t$
(let us drop the $+$ subindex for the interior quantities and $r$ when not necessary)
impose first the finite expansions for $\Epp$ and $\Ppp$ in (\ref{Pressure_epsilon})
\begin{equation}
\Epp(r,\theta)=\Eppz(r)+\Eppt(r) P_2(\cos\theta),\quad \Ppp(r,\theta)=\Pppz(r)+\Pppt(r) P_2(\cos\theta).
\end{equation}
As known, the problems for $l=0$ and $l=2$ separate. Here we consider only the $l=0$ problem.
The field equations for the sector $l=0$ provide the expressions for $\Eppz$ and $\Pppz$ plus
an equation in the form $h_0''=F(h_0',m_0',m_0)$.
A convenient auxiliary redefinition of the second order pressure is given by
$ \Psppz \defi \Pppz / 2(\Eb+\Pb) $.
On the other hand,
a barotropic EOS requires $\Eb'\Pppz-\Pb'\Eppz=0$ (for $l=0$), which, combined with the equation for
$\Eppz$ yields the equation for $m'_0$ given by (\ref{eq:m0}) renaming $\pstarHz\to \Psppz$.
Using this and differentiating the expression for $\Pppz$,
the equation for $h_0''$ can be rewritten as an equation for $\Psppz'$, which just reads
(\ref{eq:Ppf0}) with $\pstarHz\to \Psppz$.
In short, \emph{the system for $\{\Psppz(r_+),m_0(r_+)\}$ in $r_+\in(0,\ro]$
corresponds indeed to the system for $\{\pstarHz(R),m_0(R)\}$ in $R\in(0,\ro]$ given in \cite{Hartle1967},
as expected}. Equation (\ref{eq:hydro}) is a first integral of the system and determines $h_0$.
Clearly, the solutions for the exterior $m^-_0(r_-)$ and $h^-_0(r_-)$ are
given by (\ref{sol:m0ext}) with $r\to r_-$. 

The boundary conditions for these functions in $\Sigma_0$ are obtained
applying the perturbed matching theory reviewed in Section \ref{section:method}.
The whole procedure is presented elsewhere \cite{ReinaVera2014}.
In this contribution we only present one of the results. The second order matching conditions
imply that 
$m_0^\pm$ do not agree in $\Sigma_0$ in general, since they must
satisfy $[m_0] = -4 \pi \frac{a^3}{M}E(a)\mathcal{P}_0(a)$.
This fact contradicts the implicit assumption
on the ``continuity'' of $m_0$, in particular, made in \cite{Hartle1967}
(and other works, e.g. \cite{Chandra_Miller1974}, \cite{ColpiMiller}).
As a result, using the exterior solution (\ref{sol:m0ext}) for $m_0^-$
the change in mass reads, in terms of the functions used in \cite{Hartle1967}, as
\begin{equation}
  \label{eq:deltaM}
  \delta M=m_0(a)+\frac{J^2}{\ro^3}+4\pi\frac{\ro^3}{M}(a-2M)\Eb(\ro) \pstarHz(\ro),
\end{equation}
where $m_0(\ro)(=m_0^+(\ro))$ and $\pstarHz(\ro)$ are obtained from the interior problem. A jump in a ``$m_0$'' function had been found already in \cite{Bradley_etal2007}, but the discrepancy with \cite{Hartle1967} was not  established at the moment.
If the energy density of the static star $\Eb$ vanishes at the boundary
expression (\ref{eq:deltaM}) agrees with that found in  \cite{Hartle1967}, i.e. (\ref{deltaM_Hartle}).
Indeed, most of the models for neutron stars considered in the literature rely
on EOS's for which the energy density vanishes whenever the pressure does.
However, if $\Eb$ exhibits a non zero value at the boundary,
one must take into account the third
term, which may become important depending on the EOS considered.
As an example, the aforementioned models for strange quark stars which exhibit
a non zero value of $\Eb$ at the boundary \cite{ColpiMiller}, as well as those in \cite{Chandra_Miller1974} are under study at present.

\ack 
We thank Marc Mars for his most valuable suggestions.
We acknowledge financial support from projects IT592-13 (GIC12/66)
of the Basque Government, FIS2010-15492 from the MICINN, and
UPV/EHU under program UFI 11/55.
BR acknowledges financial support from the Basque Government grant BFI-2011-
250.
\section*{References}
\bibliographystyle{iopart-num}
\bibliography{references}
\end{document}